\documentclass[pre,preprint,groupedaddress, showkeys,showpacs]{revtex4}

\usepackage{graphicx}
\begin{document}


\title{The facilitated movement of inertial Brownian motors driven by a load under an asymmetric potential }


\author{Bao-quan  Ai$^{1}$}\email[Email: ]{aibq@hotmail.com}
\author{Liang-gang Liu$^{2}$}

\affiliation{$^{1}$ Institute for Condensed Matter Physics, School
of Physics and Telecommunication Engineering Laboratory of Photonic
Information Technology, South China
Normal University, 510006 Guangzhou, China.\\
$^{2}$ Faculty of Information Technology , Macau University of
Science and Technology, Macao.}


\date{\today}
\begin{abstract}
\indent Based on the recent work[ Phys. Rev. Lett. 98,
040601(2007)], we extend the study of inertial Brownian motors to
case of an asymmetric potential. It is found that some novel
transport phenomena appear in the presence of an asymmetric
potential. Within tailored parameter regimes, there exists two
optimal values of the load at which the mean velocity takes its
maximum, which means that a load can facilitate the transport in
the two parameter regimes. In addition, the phenomenon of multiple
current reversals can be observed when the load is increased.
\end{abstract}

\pacs{05. 60. -k, 05. 45. -a, 74. 25. Fy }



\maketitle

\indent


\indent Recently, Brownian motors have attracted considerable
attention simulated by research on molecular motors. These systems
can be modeled, for instance, by considering Brownian particles in
a periodic asymmetric potential and acted upon by an external
time-dependent force of zero average \cite{1,2,3}. Typical
examples are rocking ratchets \cite{4}, flashing ratchets
\cite{5}, diffusion ratchets \cite{6}, correlation ratchets
\cite{7}, white-shot-noise ratchets \cite{8}, and entropic
ratchets \cite{9,10}.

\indent Most studies have referred to consideration of the
overdamped case in which inertial term due to the finite mass of the
particle is neglected. However, there are some pervious works on
inertial Brownian motors and some novel transport phenomena were
observed \cite{11}. In particular, Machura and co-workers \cite{12}
studied the transport of inertial Brownian particles moving in a
symmetric periodic potential under the influence of both a time
periodic and a constant, biasing driving force. They found that
thermal equilibrium fluctuations can induce the phenomenon of
absolute negative mobility. Absolute negative mobility is the rather
surprising opposite behavior in the form of a permanent motion
against a (not too large) static force of whatever direction.
Haljas and co-workers \cite{13} had found that an interplay of
three-level colored and thermal noises can also generate the
phenomenon of absolute negative mobility.  In this Brief Report, we
extend the previous work \cite{12} to case of an asymmetric
potential.  We emphasize on finding whether a load can facilitate
the transport of inertial Brownian particles.

\indent Let us consider the one-dimensional of inertial Brownian
particles driving by a time-dependent external force and a load,
under the influence of an asymmetric periodic potential.  The
particle is driven by an unbiased time-periodic monochromatic force
of strength $A$ and an angular frequency $\Omega$. The equation of
motion reads as \cite{12,14}
\begin{equation}\label{}
    m\ddot{x}+\gamma \dot{x}=-V^{'}(x)+A\cos(\Omega t)+F+\sqrt{2\gamma
    k_{B} T}\xi(t),
\end{equation}
where the dot and prime denote derivatives with respect to $t$ and
$x$ respectively. The parameter $\gamma$ denotes the friction
coefficient, $T$ is temperature, and $k_{B}$ is Boltzmann constant.
$V(x)$ is the external asymmetric periodic potential and has the
period $L$ and a barrier height $\Delta V$. The quantity $F$ denotes
the external, constant force. The thermal fluctuations due to the
coupling of the particle with the environment are modeled by a
zero-mean, Gaussian white noise $\xi(t)$ with autocorrelation
function $<\xi(t)\xi(s)>=\delta(t-s)$ satisfying Einstein's
fluctuation-dissipation relation.

\indent Upon introducing characteristic length scale and time scale
\cite{12,14}, Eq. (1) can be rewritten in dimensionless form
\begin{equation}\label{}
\ddot{\hat{x}}+\hat{\gamma}\dot{\hat{x}}=-\hat{V^{'}}(\hat{x})+a\cos(\omega
\hat{t})+f_{0}+\sqrt{2\hat{\gamma}D}\hat{\xi}(\hat{t}),
\end{equation}
with
\begin{equation}\label{}
    \hat{x}=\frac{x}{L}, \hat{t}=\frac{t}{\tau_{0}},\tau_{0}^{2}=\frac{mL^{2}}{\Delta
    V}, \hat{\gamma}=\frac{\gamma \tau_{0}}{m},
    \hat{V(\hat{x})}=\frac{V(x)}{\Delta V}, a=\frac{AL}{\Delta V}, \omega=\Omega\tau_{0}
    ,D=\frac{k_{B}T}{\Delta V}, f_{0}=\frac{FL}{\Delta V},
    \end{equation}
and the zero-mean white noise $\hat{\xi}(\hat{t})$ has
auto-correlation function
$<\hat{\xi}(\hat{t})\hat{\xi}(\hat{s})>=\delta(\hat{t}-\hat{s})$.

 \indent In the following, mostly for the sake of simplicity,
we shall only use dimensionless variables and shall omit the "hat"
notion in all quantities. In addition, our emphasis is on finding
whether a load can facilitate the transport, so we take a new
quantity $f=-f_{0}$. For the asymmetric ratchet potential $V(x)$, we
choose
\begin{equation}\label{}
    V(x)=\sin(2\pi x)+\frac{\Delta}{4}\sin(4\pi x),
\end{equation}
where $\Delta$ is the asymmetry parameter of the potential. We
restrict the discussion here to a set of optimal driving parameters
\cite{12}, reading $a=4.2$, $\omega=4.9$, and $\gamma=0.9$.

  \indent Our emphasis is on finding the asymptotic mean velocity which is defined as the average of the velocity over
the time and thermal fluctuations. The Fokker-Planck equation
corresponding to Eq. (2) cannot be analytically solved, therefore,
we carried out extensive numerical simulations \cite{14}. We have
numerically integrated Eq. (2) by the Stochastic Runge Kutta method
of the second order with time step $\Delta t=0.001$. The initial
condition of $x(t)$ is taken from a uniform distribution over the
dimensionless periodic $L=1$ of the ratchet potential and the
initial condition of $v(t)$ is chosen at random from a symmetric,
uniform distribution over the interval $[-0.2, 0.2]$. The data
obtained were average over 500 different trajectories and each
trajectory evolved over $5\times 10^{4}$ periods. The numerical
results were shown in figures 1 and 2.

\begin{figure}[htbp]
  \begin{center}\includegraphics[width=10cm,height=8cm]{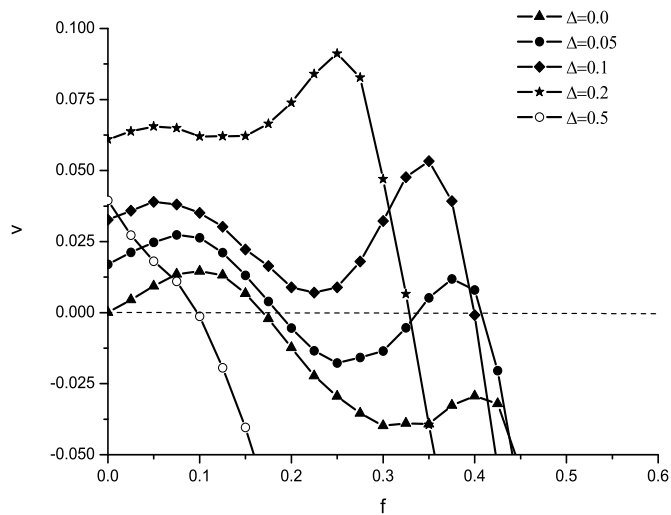}
  \caption{The asymptotic mean velocity $v$ vs the external force $f$ for different values of the asymmetry parameter $\Delta$ at $a=4.2$, $\omega=4.9$, $\gamma=0.9$, and $D=0.001$. }\label{1}
\end{center}
\end{figure}
\indent Figure 1 shows the asymptotic mean velocity $v$ as a
function of the external force $f$ for different values of the
asymmetry parameter $\Delta$.  When the potential is symmetric
($\Delta=0$), there are two peaks in the curve, one above the zero
velocity line and the other below the zero velocity line. The peak
above the zero velocity line (the left peak) is already investigated
by Machura and co-workers \cite{12}, which shows the phenomenon of
absolute negative mobility. The peak below zero velocity line (the
right peak) is induced by the high-frequency driving (the
time-dependent periodic external force) ratchet effect. When the
asymmetry parameter is increased, the position of the right peak
moves to small value of the load $f$ and its height increases. The
two peaks may be above the zero velocity line
($\Delta=0.05,0.1,0.2$), which shows that a load can facilitate the
transport in the two different regimes. However, for large value of
the asymmetry parameter ($\Delta=0.5$), the two peaks disappear and
the velocity decreases monotonically with increasing the load $f$.
It is obvious that in this case both the absolute negative mobility
effect and the high-frequency driving ratchet effect disappear and
the transport is dominated by the noise driving ratchet effect. In
addition, we also find that multiple current reversals occur as the
load is increased. The Brownian particle changes its direction three
times for $\Delta=0.05$. It must be pointed out that multiple
current reversals can be detected in overdamped Brownian systems
driven by the colored three-level Markovian noises \cite{15}.

\begin{figure}[htbp]
  \begin{center}\includegraphics[width=10cm,height=8cm]{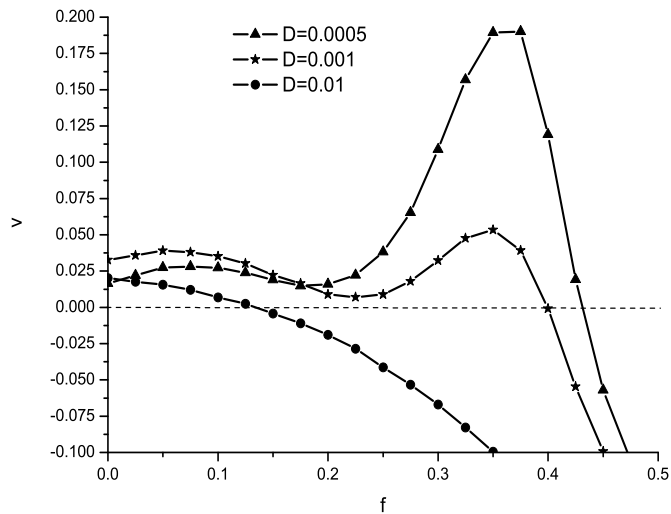}
 \caption{The asymptotic mean velocity $v$ vs the external force $f$ for different values of $D$ at $a=4.2$, $\omega=4.9$, $\gamma=0.9$, and $\Delta=0.1$.}\label{1}
\end{center}
\end{figure}

\indent In figure 2, we depict the asymptotic mean velocity as a
function of the external force $f$ for an asymmetric potential
($\Delta=0.1$). As the noise intensity is increased, the
high-frequency driving ratchet effect decreases and the noise
driving ratchet effect increases. Thus, the height of the right peak
decreases. For very large values of $D$, both the absolute negative
mobility effect and the high-frequency driving ratchet effect
disappears and the two peaks disappear.

  \indent In conclusion, in this Brief Report we study the transport of inertial Browinan particles moving
  in an asymmetric potential. Within tailored parameter regimes
  given in Ref. 12, we extend their work to the case of an asymmetric
  potential. It is found that there are two optimal values of
  the load, at which the mean velocity takes its maximum, which
  indicates that the load facilitate the transport in the two
  different regimes. The first optimal value (the left peak) is
  from absolute negative mobility \cite{12} and the second one is
  induced by the high-frequency driving ratchet effect. When the
  noise intensity $D$ or the asymmetry parameter $\Delta$ is very
  large, the two optimal values disappear. In addition, when the load is increased,
  multiple current reversals may occur within tailored parameter regimes. For example, for
  $\Delta=0.05$, the particle can change its direction three times
  as the load is increased.

\section{ACKNOWLEDGMENTS}
The work was supported by the National Natural Science Foundation of
China under Grant No. 30600122 and  GuangDong Provincial Natural
Science Foundation under Grant No. 06025073.

\end{document}